# Orientation refinement for nanobeam precession electron diffraction patterns using spot intensity profiles

Arthur Després*, Muriel Veron, Edgar F. Rauch
Univ. Grenoble Alpes, CNRS, Grenoble INP, SIMaP, F-38000 Grenoble, France
* corresponding author: arthur.despres@grenooble-inp.fr

We develop a method to measure crystallographic orientations with high angular resolution for nanobeam precession electron diffraction patterns, and applicable to 4D scanning precession electron diffraction. We show that internal variations of intensity in the diffraction spots can be simulated and correlated to experimental variations to measure the crystallographic orientation with a much better angular resolution that with the standard method of automated crystal orientation mapping (ACOM). In a tilt series carried out with long exposure time on a monocrystal of silicon around the main goniometer tilt axis, the imposed tilt angles are retrieved with a 0.044° to 0.054° precision with refinement, against 0.25° to 0.33° with the standard method. Shorter exposure times, introducing Poisson noise, marginally degrades the precision to a 0.054° to 0.066° range. The method also corrects a systematic error of ACOM near zone axis.

## 1. Introduction

Orientation mapping in the transmission electron microscope is used in many fields of materials science, e.g. microelectronics, metallurgy, proteins, biomaterials, where structural information at high spatial resolution is required. The standard method for orientation mapping is automated crystallographic orientation mapping (ACOM) [1], replicated with some variations in the data processing elsewhere [2,3]. To achieve both high spatial resolution and angular resolution, the acquisition is carried out with a nanobeam of a few mrad in convergence semi-angle, coupled with precession, resulting in diffraction spots with a finite radius. When precession is involved, as in the present work, the acquisition setup is referred to as 4D scanning precession electron diffraction (4D-SPED). Then experimental patterns are correlated to theoretical patterns, also called templates, calculated assuming kinematical diffraction conditions. The angular precision of the measure usually reaches 0.5° with the standard processing algorithm, and 0.12° with a more advanced processing algorithm taking account of the effect of precession on the intensity of the diffraction spots [4]. While such an angular precision is more than sufficient for characterizing grain boundaries and texture, it remains limited for small misorientations at defects such as dislocations, subgrain boundaries and inclusions.

More recently, another kind of method has been proposed by Palatinus *et al.* for treating precession electron diffraction patterns [5]. The method essentially uses a physically more accurate description of the intensity and position of diffraction spots to refine the calculated orientation. It is very well adapted to acquisitions with large illuminated area and parallel beam, where it reaches an angular precision of 0.018° to 0.05°. On the contrary, it is not really designed for nanobeam precession electron diffraction, as it considers punctual diffraction spots approximated by Gaussian intensity profiles.

Correa *et al.* have proposed an extension of the model of Palatinus to specifically treat nanobeam precession electron diffraction patterns [6]. The method considers many physical effects, but no particular treatment is given to the finite size of the spots (they are simulated as disks, but this is not presented as a feature of interest). The results are not entirely satisfactory, as they fail to retrieve the expected rotation field in the test case (screw dislocation in an InP nanowire).

The reason why the finite size of the diffraction spots is important in nanobeam precession electron diffraction patterns is because some diffraction spots may present internal variations of intensity. This phenomenon is largely ignored in current processing methods of orientation mapping. However, it is ackowledged in methods of strain mapping [7–9], where the position of the diffraction spots must be located with high precision and accuracy. In strain mapping several strategies have been designed to cancel the variations of intensity that are naturally present in the diffraction spots, but none of them takes advantage of these variations to improve the measurement.

In this paper, we will show that the internal variations of intensity in nanobeam precession electron diffraction patterns can easily be modelled and used to refine orientation measurement. A tilt series carried out on silicon will serve as test case, and the refinement method will be compared to the standard ACOM method. Finally its applicability to orientation mapping and its potential for improving strain mapping will be discussed.

## 2. Method

### 2.1. Material and acquisition

The material is a silicon single crystal taken from microelectronics wafer. The sample was prepared by cross-section polishing following standard procedures, with a $\langle 110 \rangle$ direction normal the surface. Observations were conducted in a JEOL 2100F TEM equipped with a field emission gun operating at 200kV. Diffraction patterns were acquired with a focused beam of 1.2mrad convergence angle, and the precession angle was set to 1°. Precession was carried-out with the ScanDpa software from Nanomegas. The diffraction patterns were recorded with a 512x512 pixels in-line direct electron detector Medipix 3 implemented in the Cheetah camera from Amsterdam Scientific Instrument. The exposure time per pattern was set to 10s to minimize Poisson noise and improve detection of the high Miller indices and low intensity diffraction spots. To be more exact, each pattern is a sum of 1000 patterns acquired with a 10ms exposure time, but this has no consequence for the resulting Poisson noise. The influence of the exposure time will be considered in the discussion section. Finally, in order to remove the continuous background caused by inelastic scattering, a median filter with kernel width of 35 pixels was applied during post-processing. The kernel width is several times larger than the diffraction spots size to ensure that the spots intensity are not affected by the filter. The sample was tilted around the goniometer tilt axis by 1° steps from the nominal value of -18° to 12°. For ease of reading, we will note them 0° to 30° in the rest of the paper. The tilt axis is at 62° from the recording camera x axis.

### 2.2. Examination of acquired diffraction patterns

Two examples of diffraction patterns acquired during the tilt series are shown in Figure 1. The pattern in Figure 1a was taken in 110 zone axis, while that in Figure 1b is tilted of 6° from the zone axis. The diffraction spots radius being in theory proportional to the convergence angle, the diffraction spots should appear circular. However, some diffraction spots exhibit strong intensity gradients within their radius, which gives them a more lenticular shape. The $\bar{1}91$, $64\bar{6}$ and $\bar{5}15$ spots in Figure 1a and the $\bar{5}\bar{9}7$ and $\bar{7}\bar{1}7$ spots in Figure 1b are indicated as examples but many more can be observed. The short length of these diffraction spots follows the radial direction with respect to the transmitted beam. This phenomenon is analogous to the excess lines that appear in convergent beam electron diffraction patterns. The two main differences with respect to convergent beam diffraction are that the diffraction spot radius is here small because of the small convergence angle, and that the diffraction spot intensity is affected by precession.

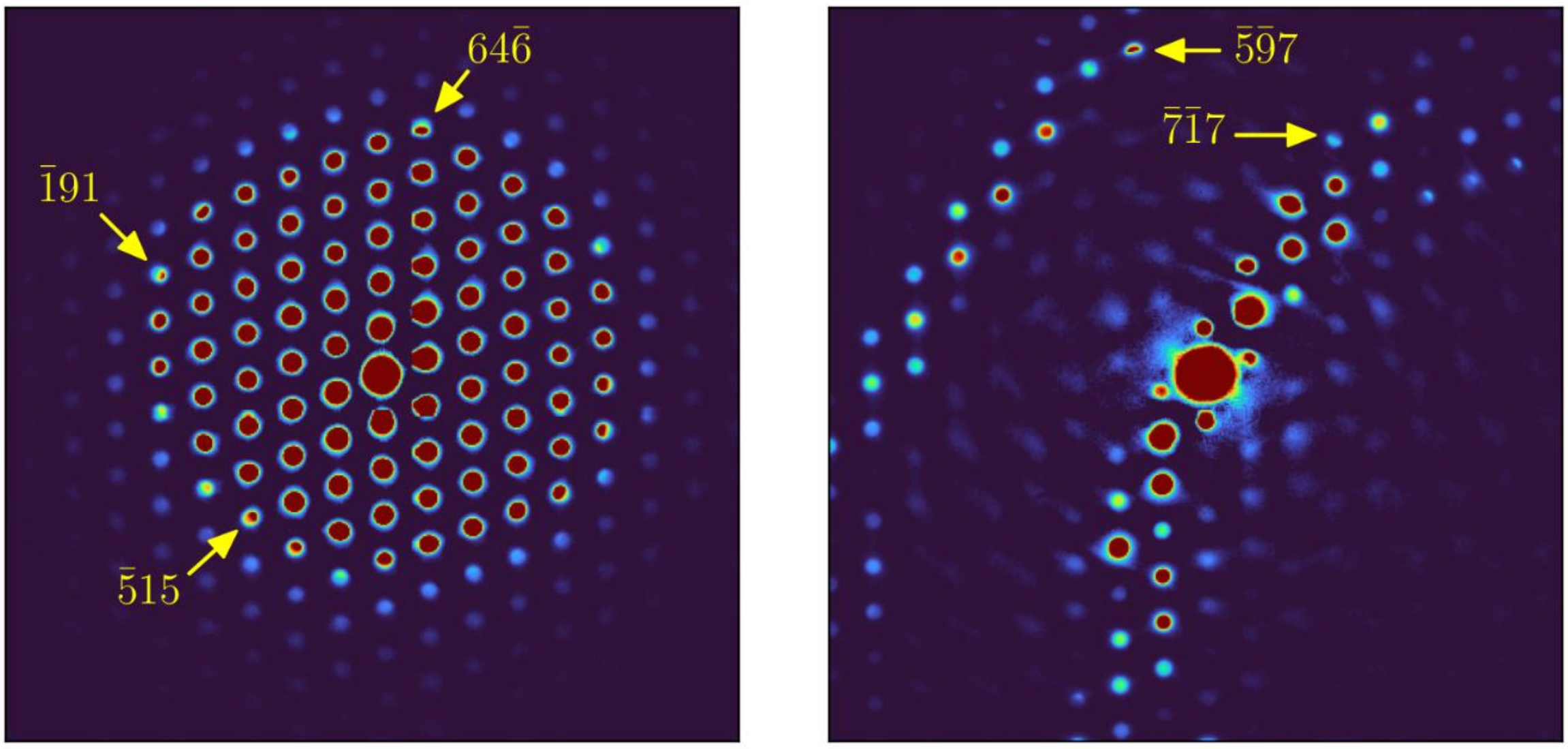

(a) (b)

Figure 1 : Diffraction patterns taken (a) in 110 zone axis, (b) away from 110 zone axis. The patterns are shown after application of the median filter. The saturation of some diffraction spots is only an effect of display.

## 2.3. Intensity distribution in a diffraction spot

We show here that internal variations of intensity in the diffraction spots are related to local changes of diffraction conditions. For this, first, we consider a *homogeneous* intensity distribution, as this is the case for most reflexions close to the Bragg condition. In a subimage centred on the diffraction spot, we define the polar coordinate system $(\alpha, \varphi)$, with $\alpha$ the radius and $\varphi$ the polar angle. We note that $\alpha$ also corresponds to the local convergence angle. The intensity distribution is described by a generalized Lorentz function. It simulates a plateau of intensity in the diffraction spot radius, and a Lorentzian decay outside [10]. Lorentzian decay being frequent in scattering and diffraction phenomena, it is here preferred to Gaussian decay. The homogeneous intensity thus follows:

$$I_{hom} = \frac{1}{(\alpha/\alpha_0)^n + 1} \quad (1)$$

Where $\alpha_0$ is the diffraction spot radius, equivalent to the nominal convergence angle, and $n$ controls the steepness of the intensity decay outside the diffraction spot. These parameters are fitted on one of the most intense diffracted spots, and kept constant for the whole pattern.

Secondly, we consider the *heterogeneous* intensity distribution induced by variations of excitation error with the local convergence angle and with precession. At a given position $(\alpha, \varphi)$, in the small angle approximation, the excitation error in the radius of an $hkl$ diffraction spot is given by:

$$s = s_g + g\alpha cos(\varphi - \varphi_0) \quad (2)$$

Where $s_g$ is the excitation error estimated for this reflexion, $g$ is the diffraction vector norm, and $\varphi_0$ is the polar angle between the transmitted beam and the diffraction spot. This formula expresses the change of excitation error as the Ewald sphere is tilted due to changes in local convergence angle [5]. For the demonstration, we assume here that $s_g$ is precisely known.

Palatinus *et al.* [5] have proposed to assume two beam conditions to calculate the intensity with precession. Even though this is a crude approximation of the real conditions, we keep it for nanobeam patterns as it has demonstrated its effectiveness for indexing parallel beam diffraction patterns. After a few mathematical developments recalled in Appendix A, the intensity of a reflexion when precession is activated follows:

$$I_{pr} = \int_0^{2\pi} \frac{1}{\xi^2\left(s + \beta g cos(\psi)\right)^2 + 1} d\psi \quad (3)$$

Where $\xi$ is from a rigorous point of view the extinction distance, $\beta$ is the precession angle, and $\psi$ the precession phase. However, for large extinctions distances, corresponding to high Miller indices reflexions, it is recommended to set $\xi$ as a constant of the order of magnitude of the foil thickness in order to better fit the overall shape of the intensity curve [5]. The integral has an analytical solution expressed by:

$$I_{pr} = \frac{2\pi}{z}\sqrt{2(u + z)} \quad (4)$$

With $u = \xi^2((g\beta)^2 - s^2)$ and $z = (u^2 + 4(\xi s)^2)^{1/2}$. The intensity as a function of excitation error when precession is activated follows a so-called camel plot, with two maximums of intensity at $+g\omega$ and $-g\omega$.

Finally, to account for the decay of intensity outside the diffraction spot, the heterogeneous intensity distribution is masked by the homogeneous distribution:

$$I_{het} = I_{pr} \cdot I_{hom} \quad (5)$$

Where $\cdot$ is the element-wise matrix product. On the contrary to the methods of Palatinus *et al.* [5] and Correa *et al.* [6], the model presented here is not concerned with predicting intensity variations from one diffraction spot to the other but the internal variations within a given spot. Reflexions associated with excitation errors near $+g\omega$ or $-g\omega$ will be the most likely to exhibit large variations.

For simplicity, $\xi$ was taken as a constant parameter for all diffraction spots and all patterns. After a parametric study aiming to minimize the correlation error between simulated patterns and experiments, it was set to 500Å. This is a realistic value for a foil thickness.

Figure 2 shows the simulated intensities and their comparison to the experiment for the $\bar{1}91$, $64\bar{6}$ and $22\bar{2}$ diffraction spots in the 110 zone axis diffraction pattern. The intensity distributions were predicted after the orientation refinement procedure described in the next section, meaning that the orientation is here known precisely. Figure 2a shows the homogeneous intensity distribution for the $\bar{1}91$ spot (identical for all diffraction spots), and illustrates the polar coordinate system. Figure 2b shows the precession intensity distribution calculated for this spot, with a line of maximum intensity crossing the field of view. Figure 2c shows the resulting heterogeneous intensity distribution for this spot. There is an excellent agreement with the experimental distribution shown in Figure 2d.The same comparison is shown in Figure 2e to h for the $64\bar{6}$ spot, with an excellent agreement as well between the heterogeneous and experimental distributions. The main difference is that the short length of the diffraction spot has been rotated, as a result of the different polar angle $\varphi_0$ with respect to the transmitted beam. Finally, Figure 2i to l show the same comparison for the $22\bar{2}$ spot. Here the intensity distribution in the experimental spot is quasi homogeneous. Because the precession intensity distribution is homogeneous for this reflexion, ultimately the heterogeneous intensity distribution remains homogenous, in agreement with the experiment.

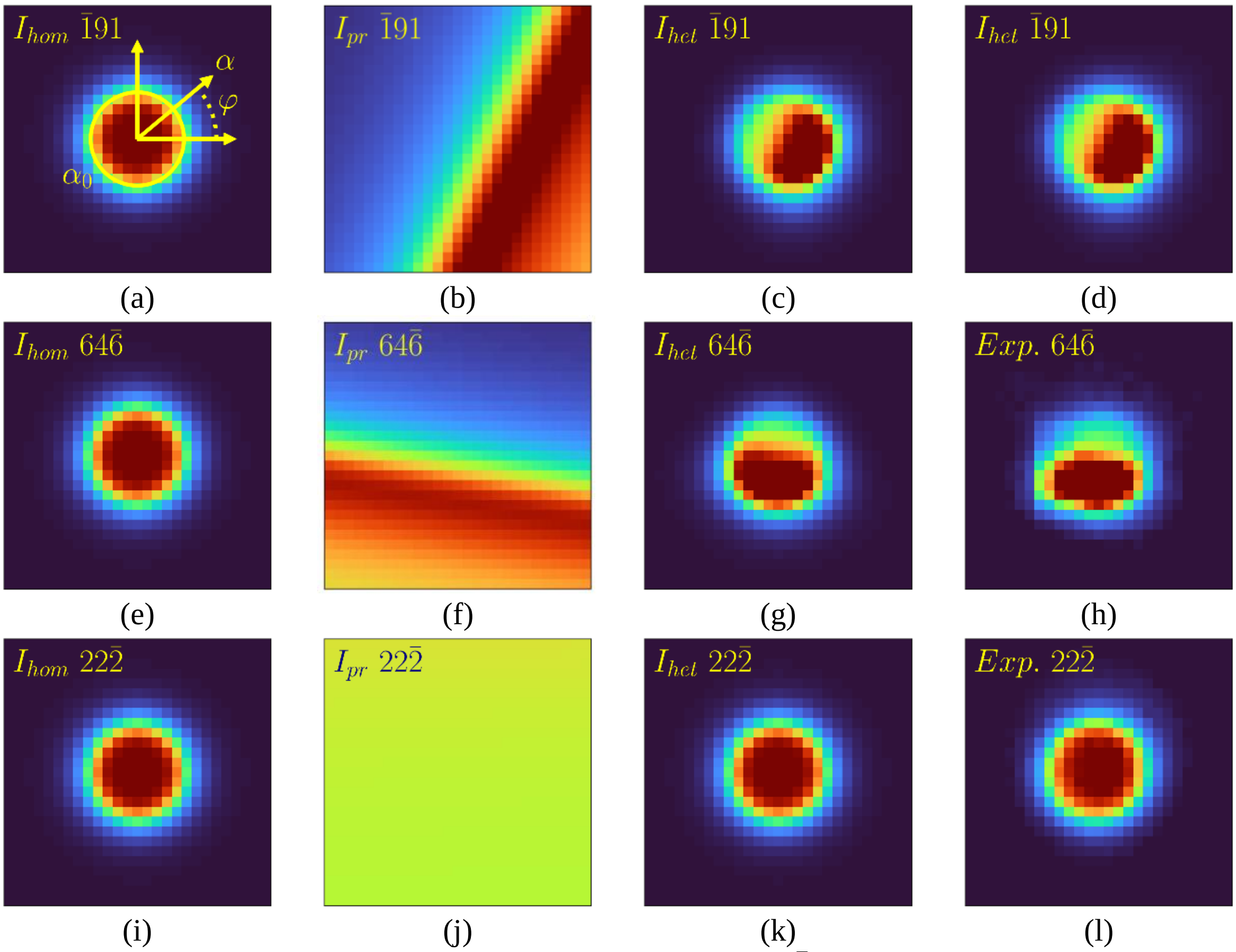

Figure 2 : (a-d) Simulated intensities and comparison to the experiment for the $\bar{1}91$ diffraction spot in the 110 zone axis diffraction pattern.(e-h) A second example for the $64\bar{6}$ spot. (i-j) A third example for the $22\bar{2}$ spot. The images are normalized to show the same average intensity.

### 2.4. Procedure for orientation refinement

For a given orientation, the theoretical positions of the diffraction spots $\boldsymbol{g}$ and their excitation errors $s_g$ are calculated by projecting the intersection point between the Ewald sphere and the reciprocal lattice rods on the camera plane. From the excitation errors and theoretical positions, the heterogeneous intensity distributions $I_{het}$ of the diffraction spots are also calculated.

The agreement between the theoretical and the experimental spots is measured by phase cross correlation. This method operates in the Fourier space, and provides for each diffraction spot a shift in position (converted into an experimental position $\boldsymbol{g_{exp}}$), and a normalized root mean square error. In the particular context of phase cross correlation, this error metric is invariant to shift and to the scale of intensity, as demonstrated in earlier works [11,12].

The orientation refinement is carried-out in three steps. First, an initial orientation is calculated by ACOM [1] to serve as initial guess. Secondly, the out-of-plane component of the orientation is refined. The initial orientation is rotated out-of-plane in order to minimize the mean of all diffraction spots errors. More details on the minimization are provided in Appendix B. Thirdly, the in-plane component of the orientation is refined. The rotation between the theoretical and experimental and theoretical spot positions is calculated from the deformation gradient tensor $\boldsymbol{F}$ that transforms the theoretical spot positions $\boldsymbol{g}$ to the experimental spot positions $\boldsymbol{g_{exp}}$. The in-plane rotation arises as the diagonal component of the antisymmetric part of the deformation gradient tensor $(\boldsymbol{F} - \boldsymbol{F^T})/2$.

The distinction between out-of-plane and in-plane components of the orientation is necessary because out-of-plane rotations induce mostly intensity variations in the diffraction spots, while in-plane rotations only shift the diffraction spot positions. It must also be pointed out that the in-plane rotations are insensitive to the elliptic distorsions traditionally observed on diffraction patterns [13,14]. This is because elliptic distorsions affect the symmetric part of the deformation gradient tensor (equivalent to a plane strain deformation of the pattern), while in-plane rotations are included in the antisymmetric part.

## 3. Results

Figure 3 compares the orientations calculated with ACOM and with refinement during the tilt series. The orientations are plotted in 001 pole figures. In Figure 3a, the orientations calculated with ACOM show some scattering, which appears more clearly in the inset. In addition, there are two gaps with no indexed orientations in the tilt series. These gaps correspond to orientations close to the zone axis orientation, which the ACOM misses by several tenth of degrees. This error of the ACOM indexation near zone axes is known by the users of this method [6]. By comparison, the orientations calculated with refinement, in Figure 3b, show much less scattering along the tilt series, and there are no missing gaps near the zone axis.

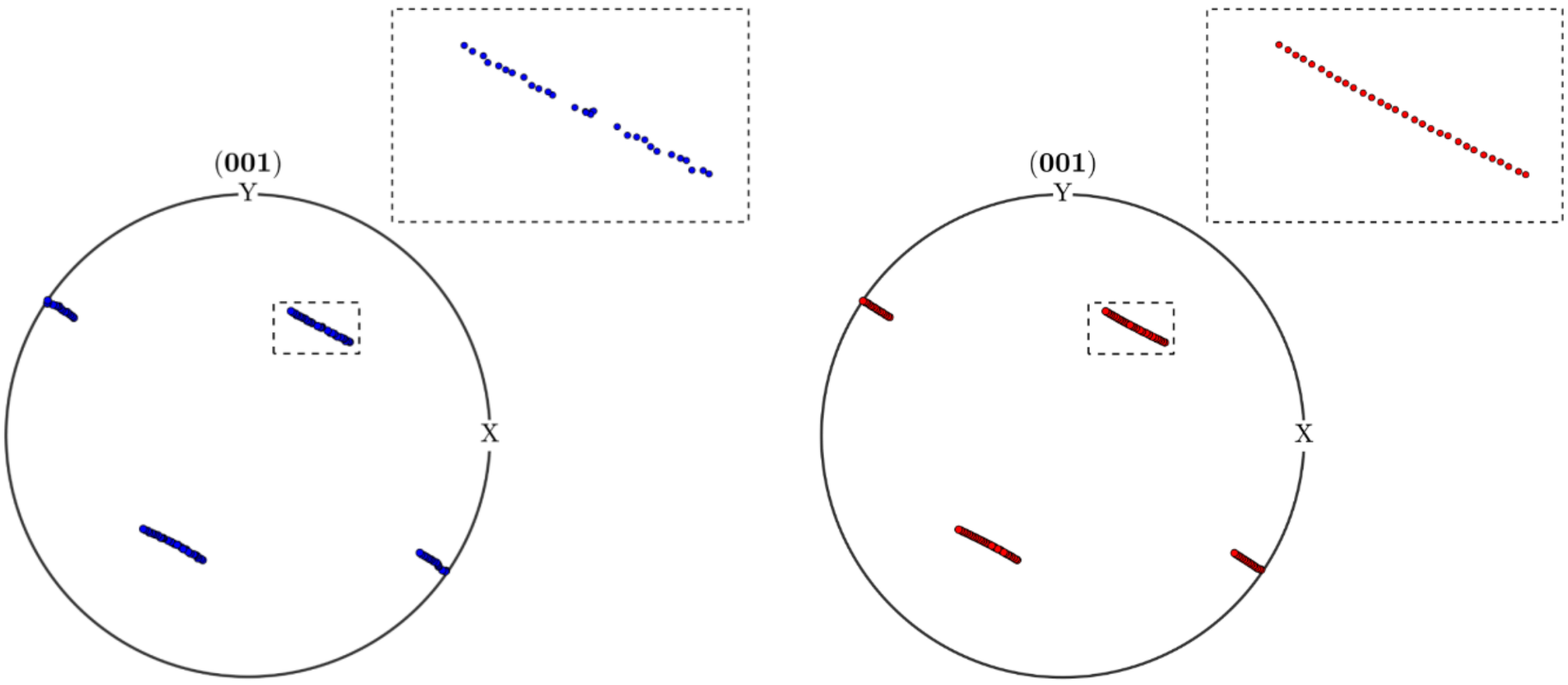

(a) (b)

Figure 3: 001 pole figures of the orientation measured during the tilt series with (a) ACOM, (b) orientation refinement. Zooms on one 001 pole is provided in the insets.

In electron microscopy, the angular resolution is usually quantified from the point-to-origin misorientation angles in a tilt series [5,15,16]. The misorientation axis-angle **r** calculated between the first orientation and each orientation in the tilt series is decomposed into its three directional components $r_x$, $r_y$ and $r_z$. The two first components make the out-of-plane rotation, while the third term is the in-plane rotation. In Figure 4a, the point-to-origin misorientation angles calculated from the ACOM orientations are plotted as a function of the nominal tilt imposed to the goniometer. The precision of the measure is defined as the root mean square error between the experimental points and their linear regression. It ranges between 0.276° and 0.339° depending on the component. These values are higher than the previously reported precision of 0.12° for ACOM [15]. This is explained by the systematic error close to the zone axis, already observed in Figure 3a, and manifested, in this Figure 4 a, as an oscillation of the misorientation angles around the linear trend in the 10° to 20° range. If these values are excluded, an error of 0.18° is found, which is closer to the state-of-the-art. By comparison, Figure 4b shows the misorientation angles calculated with orientation refinement. The precision now ranges from 0.044° to 0.054°, which is well below the best measurements made by ACOM [15].

By geometry, the nominal rotation rate is $\cos(62°)°$ per degree of tilt for the $r_x$ component, $\sin(62°)°$ for the $r_y$ component and 0° for the $r_z$ component. The measured rotation rate is the slope of the linear regression, and the accuracy is taken as the difference between the nominal and measured rotation rates. For the orientations measured by ACOM, the accuracy is below 0.014° for all component, which indicates that ACOM is on average an accurate method (although the accuracy will be worse if only orientations near zone axis are considered). The accuracy remains in the same range after orientation refinement.

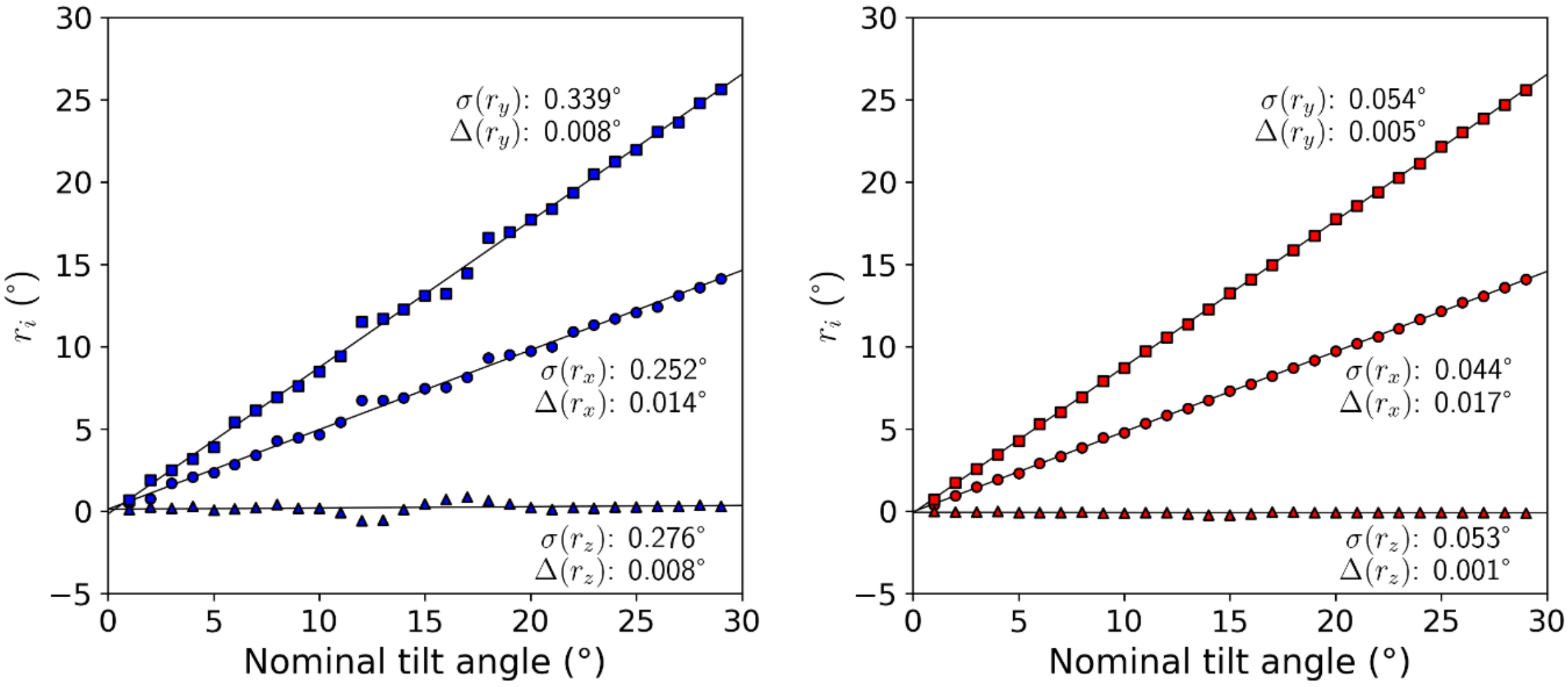

(a) (b)

Figure 4: Point-to-origin misorientation angles calculated with (a) ACOM, (b) orientation refinement. Round symbols correspond to the $r_x$ component, square symbols to $r_y$, and triangles to $r_z$.

A visually striking illustration of the improvement in angular resolution with orientation refinement is the measure of the point-to-point misorientation axis. This parameter is known to be subject to an important error for low increments of tilt angles [17]. Figure 5a shows a large scattering for the axes calculated with the ACOM orientations. The root mean square deviation from the goniometer tilt axis is 12.1°. Figure 5b shows a much reduced scattering after orientation refinement, with a mean square deviation of only 1.7°.

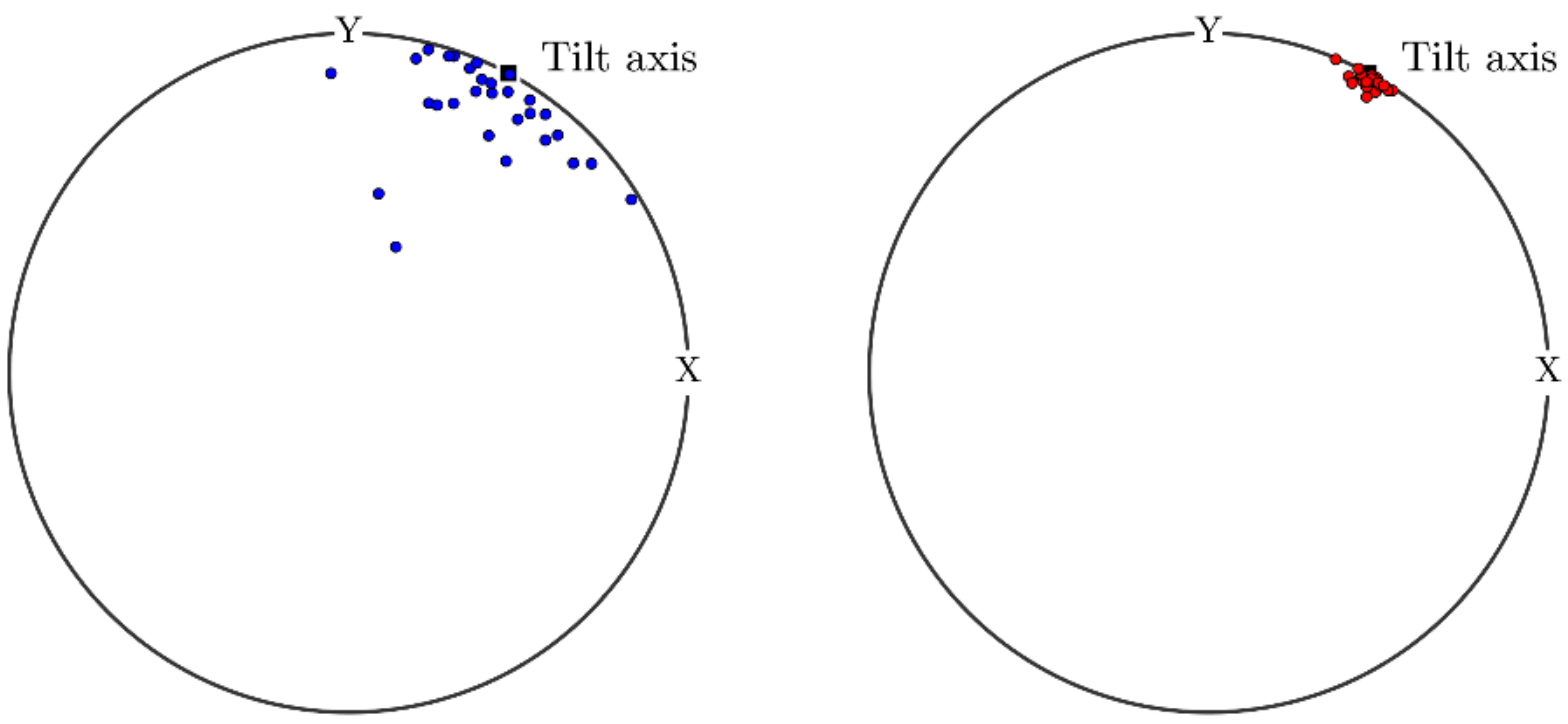

(a) (b)

Figure 5: Point-to-point misorientation axes calculated with (a) ACOM, (b) orientation refinement.

## 4. Discussion

### 4.1. Applicability to orientation mapping by 4D-SPED

The present work focused on the implementation of an orientation refinement method applicable to nano beam precession electron diffraction patterns. To minimize Poisson noise and to clearly associate internal variations of intensity in the diffraction spots to diffraction effects, we have used an effective exposure time per pattern of 10s by summing 1000 patterns of 10ms acquisition. However, for orientation mapping by 4D-SPED, where the objective is to scan rapidly large regions of interest and obtain in a reasonable time a large number of diffraction patterns, the acquisition rate is usually of the order of 10ms.

To evaluate whether orientation refinement could be carried out on patterns acquired with faster acquisition rates, the refined orientation were recalculated with an effective exposure time of 10ms (however, the simulated diffraction intensities were kept without Poisson noise). Figure 6 shows the resulting point to origin misorientation angles. The error introduced by Poisson noise is marginal, as the precision in misorientation angles increases of about 0.01° for each component, while the accuracy remains excellent. Importantly the precision remains well below that obtain with ACOM. This result could be expected because phase cross correlation algorithms are known to be very robust to noise [11]. Fast acquisition rates of the order of 10ms would not dramatically change the precision of orientation refinement. We therefore conclude that orientation refinement should be applicable to orientation mapping using 4D-SPED datasets obtained at fast acquisition rates.

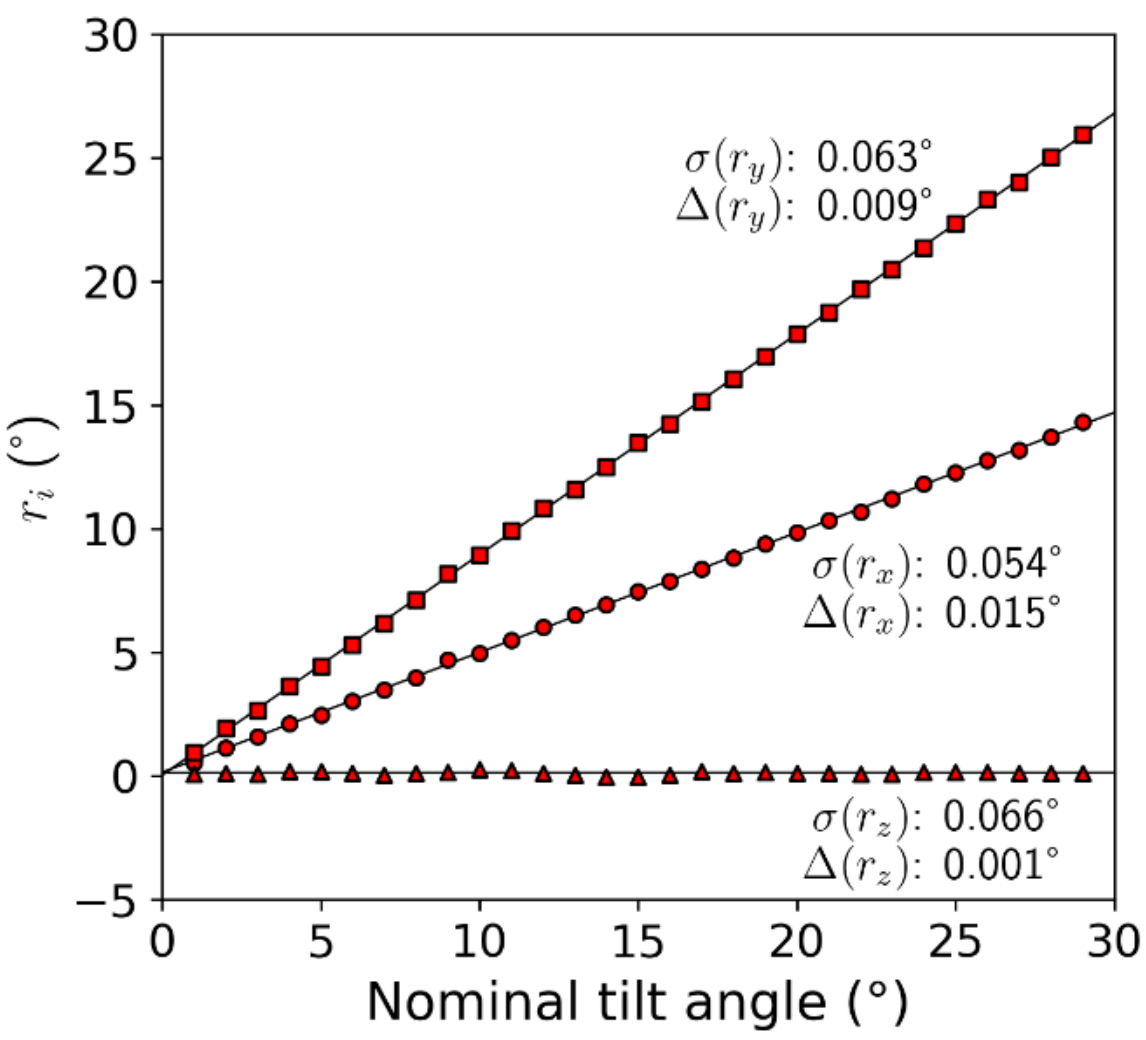

Figure 6: Point to origin misorientation angles calculated with refinement and 10ms of exposure time per pattern.

### 4.2. Interest for strain mapping

In strain mapping, lattice strains are calculated from the fine positions of diffraction spots. Since distances between diffraction spots and the transmitted beam are inversely proportional to lattice interplanar distances, differences in position of diffraction spots between patterns can be related to lattice strains [8,18–20]. To locate diffraction spots with sub-pixel accuracy, several approaches have been implemented, such as cross correlation with a homogeneous diffraction spot [9,20], centre of mass [19], and radial gradient maximization assuming circular diffraction spots [21]. In this context, internal variations of intensity are usually considered as an undesired feature, because they make the fine position of the diffraction spots more difficult to retrieve. Using precession is a successfull strategy to reduce these internal variations [22], but the present work shows that it cannot fully remove them due to physical reasons. It can be anticipated that increasing the precession angle would make diffraction spots with strong internal intensity variations appear at larger reciprocal distances, but that would also increase the number of such spots.

In future applications, internal variations of intensity could be used to better locate the fine position of diffraction spots, and consequently improve strain mapping in 4D-SPED datasets. The proposed method would parallel those of Zeltman *et al.* [9] and Mahr *et al.* [23], who have shown that taking account of the internal variations of intensity is a successfull strategy for improving the precision of strain measurement. In their method, implemented on datasets without precession, the intensity variations were artificially controlled using a patterned condenser aperture, giving to the probe and diffraction spots the shape of the patterned aperture. In the proposed method the patterned probe would be a natural result of the diffraction conditions.

## 5. Conclusion

We have developed a method of crystallographic orientation refinement for nanobeam precession electron diffraction patterns. It relies on the simulation of diffraction spot intensity profiles, and particularly their internal variations of intensity. The angular precision of the method reaches 0.044° to 0.054° for patterns with long exposure times, and 0.054° to 0.067° for shorter times. It also corrects a systematic error of ACOM near zone axis orientations. The method is robust to Poisson noise, and for this reason, has a strong potential for being applied to high resolution orientation and strain mapping with large 4D-SPED datasets.

## Ackowledgements

Lætitia from LMGP laboratory (Grenoble) is warmly thanked for provision and preparation of the silicon sample.

## CRediT authorship contribution statement

**Arthur Després**: Conceptualization, Investigation, Data curation, Writing – original draft. **Muriel Véron**: Investigation, Writing – review & editing. **Edgar F. Rauch:** Conceptualization, Writing – review & editing.

## Data availability

Data will be made available on reasonable request.

## Appendix

### Appendix A: Intensity of diffracted beam with precession

A simple and reasonably accurate expression of the intensity of a diffracted beam as a function of the excitation error is the rocking curve for two beam conditions:

$$I = \left(\frac{\pi t}{\xi}\right)^2 \frac{\sin^2(\pi t\sqrt{s^2 + \xi^{-2}})}{\pi t\sqrt{s^2 + \xi^{-2}}} \quad \text{(A.1)}$$

Where $t$ is the sample thickness, $s$ is the excitation error, and $\xi$ the extinction distance. As the $\sin^2$ term oscillates between 0 and 1, the upper envelope of the rocking curve is given by a Lorentz function:

$$I_{up} = \frac{1}{\xi^2 s^2 + 1} \quad \text{(A.2)}$$

The two beam rocking curve and its envelope are shown in Figure A. *1*. The absolute maximum at $s = 0$ may be overestimated as it is unlikely that $\sin^2(\pi t\xi^{-1}) = 1$ for a random value of $t\xi^{-1}$. However, the rest of the envelope is correctly captured. This expression is kept for the following calculations.

When precession is activated, the excitation error varies as $s + g\omega\cos(\psi)$, with $\omega$ the precession angle and $\psi$ its phase (note that this is the same relation as for the effect of convergence angle on the excitation error). Then, the intensity with precession is calculated as the sum of the intensities over a turn of precession:

$$I_{pr} = \int_0^{2\pi} \frac{1}{\xi^2\left(s + \beta g cos(\psi)\right)^2 + 1} d\psi \quad \text{(A.3)}$$

The analytical solution is:

$$I_{pr} = \frac{2\pi}{z}\sqrt{2(u + z)} \quad \text{(A.4)}$$

With $u = \xi^2((g\beta)^2 - s^2)$ and $z = (u^2 + 4(\xi s)^2)^{1/2}$. The resulting intensity profile is also shown in Figure A. 1. It exhibits two maxima located at approximately $\pm g\omega$.

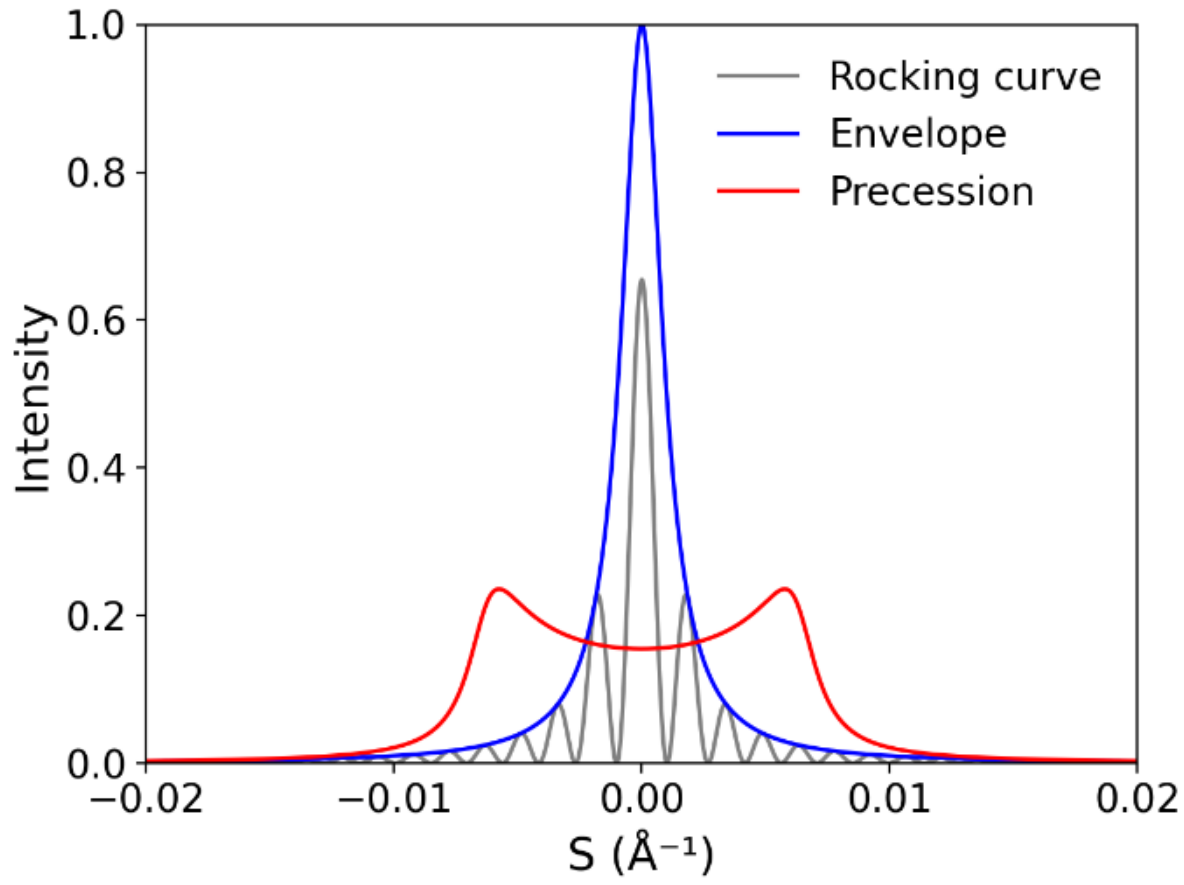

Figure A. 1 : Rocking curve, its envelope and the precession intensity for $\xi = 1000$Å, $t = 1000$Å and $\omega = 1°$. A diffraction vector length of $g = 0.37$Å$^{-1}$ is considered.

## Appendix B: Optimization of the out-of-plane orientation component

To optimize the out-of-plane orientation component, the ACOM orientation is taken as the initial guess to simulate diffraction spot intensities. This orientation can be rotated by $\mathrm{r_x}$ along the x axis and $\mathrm{r_y}$ along the y axis. Each set of $(\mathrm{r_x}, \mathrm{r_y})$ values represents an orientation with a slight out-of-plane rotation compared to the ACOM orientation. Figure B. 1(a) illustrates the error for the $\bar{1}91$ spot in the 110 zone axis pattern. The error is minimized along one of two lines corresponding to the $+\omega$ and $-\omega$ values of beam tilt caused by precession. The lines are parallel to the radial direction of the diffraction spot being investigated. Figure B. 1(b) shows the mean error field for the $\bar{1}91$, $64\bar{6}$, and $22\bar{2}$ spots. A single minimum naturally appears at the intersection of three lines corresponding to the three diffraction spots. Figure B. 1(c) shows the mean error when all spots are considered. The minimum is more clearly defined at the same intersection. It defines the out-of-plane orientation component.

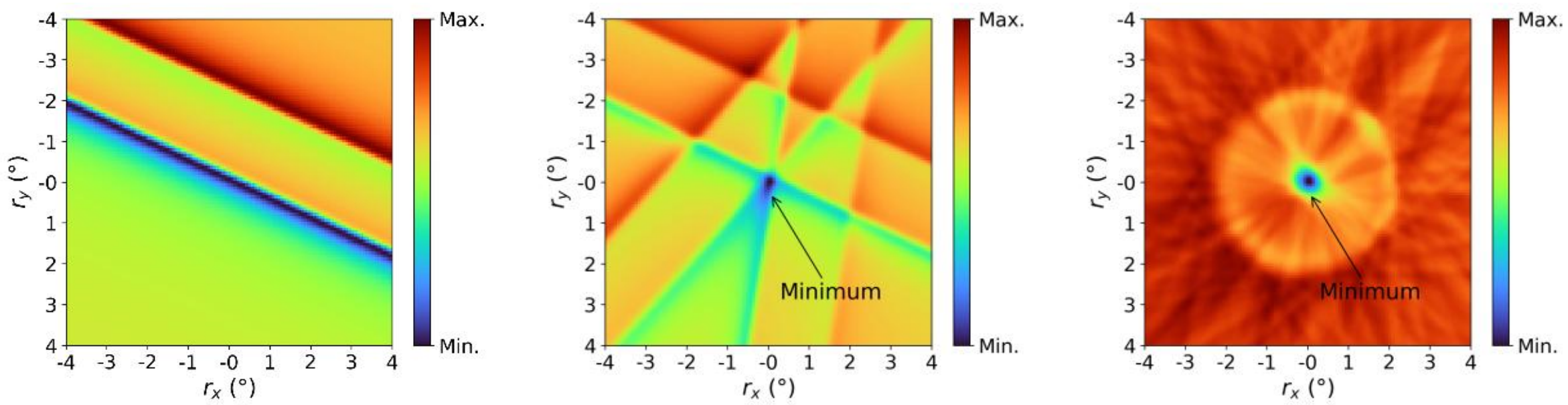

(a) (b) (c)

Figure B. 1 : Mean of the normalized root mean square errors in the $(r_x, r_y)$ space for (a) the $\bar{1}91$ spot, (b) the $\bar{1}91$, $64\bar{6}$, and $22\bar{2}$ spots, (c) all spots in the 110 zone axis diffraction pattern.